\documentclass[twocolumn,amsmath,prl,aps,showpacs,floatfix,superscriptaddress]{revtex4}

\usepackage{epsfig}
\usepackage{epsf}
\usepackage{color}
\usepackage{float}
\usepackage{graphicx}

\def\go        {\omega}
\def\gee        {\epsilon}
\def\ga        {\alpha}
\def\gb        {\beta}
\def\gc        {\gamma}

\def\qq         {{\bf q}}

\def\rr         {{\bf r}}

\renewcommand{\[}{\left[}
\renewcommand{\]}{\right]}
\renewcommand{\(}{\left(}
\renewcommand{\)}{\right)}

\restylefloat{figure}

\begin{document}

\title{Optical saturation driven by exciton confinement in molecular-chains: a TDDFT study}

\author{Daniele Varsano}
\affiliation{
National Center on nanoStructures and Biosystems at Surfaces (S3)
of INFM--CNR, Via Campi 231/A, 41100 Modena, Italy
}
\affiliation{European Theoretical Spectroscopy Facility (ETSF)}

\author{Andrea Marini}
\affiliation{Istituto Nazionale per la Fisica della Materia e
Dipartimento di Fisica dell'Universit\'a di Roma ``Tor Vergata'', 
Via della Ricerca Scientifica, I-00133 Roma, Italy}
\affiliation{European Theoretical Spectroscopy Facility (ETSF)}

\author{Angel Rubio}
\affiliation{
Dpto de F\'{\i}sica de Materiales, Universidad del Pa\'\i s Vasco, Centro Mixto CSIC-UPV, and Donostia International Physics Center, Edifio Korta, Avenida de Tolosa 72, E-20018 San Sebasti\'an, Spain}
\affiliation{European Theoretical Spectroscopy Facility (ETSF)}

\date{\today}

\begin{abstract}
We have identified excitonic confinement in one-dimensional molecular chains
(i.e. polyacetylene and H$_2$) as the main driving force for the saturation
of the chain polarizability 
as a function of the number of molecular units.
This conclusion is based on 
first principles time--dependent density functional theory 
calculations performed with a new derived exchange--correlation kernel.
The failure of simple local and
semi--local functionals is shown to be related to the lack of memory effects, spatial
ultranonlocality, and self--interaction corrections.
These effects get smaller
as the gap of the system reduces, in which case such simple approximations
do perform better. 
\end{abstract}


\maketitle

The electronic quantum confinement 
occurring in low--dimensional
systems is often responsible for the peculiar
spectroscopic properties exhibited by molecules and nanostructures.
Exciton confinement explains, for example,
the size--dependent color of semiconducting quantum
dots used as fluorescent markers in biology~\cite{Alivisatos}.
Highly localized excitons (solitons) play also a fundamental role in
describing the process of vibrational energy transfer
in complex proteins~\cite{solitons}.  In this context,
one--dimensional polymers and molecular chains constitute an
excellent playground to analyze the interplay between
correlation effects and quantum confinement. For example,
in non conducting polymers 
the longitudinal linear polarizability per monomer unit  $\alpha(N)/N$
tends to a constant in the large--N limit~\cite{resta}.
This optical saturation stems from the 
polarization of the electrons along the chain that tends to
counteract the external field and manifest itself in the dependence of the
optical absorption spectra on the polymer size. 
The present Letter aims to provide a consistent description
of this saturation within density--functional based schemes.

The commonly used local\,(LDA) and semilocal\,(GGA) approximations to
Density--Functional--Theory\,(DFT) and Time--Dependent DFT\,(TDDFT)
that successfully describe the
electronic properties of many different physical systems~\cite{tddft},
fail dramatically for the case of semiconducting
one--dimensional molecular chains. 
ALDA do not describe the main features of the absorption spectra and 
strongly overestimates the polarizability with respect to
quantum chemical calculations~\cite{toto,dalskov,champagne}.
The reason for this poor performance of ALDA is still not settled, and
it has been traced back to the  need of long-range\,(LR) terms
({\it ultra non--locality}) in the exchange-correlation functionals~\cite{gonze}
or to the lack of self--interaction correction\,(SIC)~\cite{mori,burke}.
Exact-exchange\,(EXX) or current-density functional\,(CDFT) 
approaches capture some of the effects, but not all: while
EXX works fairly well in the case of
the finite H$_2$ chain (reproducing the Hartree--Fock results~\cite{mori}) it
fails in reproducing the absorption spectrum of the infinite
trans--polyacetylene chain~\cite{rohra}.
On the contrary CDFT
provides quite good results for
finite $\pi$-conjugated polymers~\cite{vanfaassen,vanfaassen2},
while it breaks down for the  H$_2$ chain.

In TDDFT all the effects beyond the non interacting particles approximation
are embodied in the exchange--correlation\,(xc) kernel $f_{xc}\equiv\delta v_{xc}/\delta n$, with
$v_{xc}$ the xc--potential and $n$ the exact ground state electronic density.
The recent developments in  merging
many--body perturbation--theory\,(MBPT)~\cite{rmp}
and TDDFT~\cite{sottile,andrea} open the path to unravel the physical
origin of the response properties of one-dimensional systems.
Here we show that exciton confinement
dictates the  evolution (saturation) of the optical response of one--dimensional
chains as a function of the chain length. 
Therefore the failure of EXX, available
local or semilocal or CDFT approximations is related to their inability to
describe strong excitonic effects in anisotropic low-dimensional systems.
We show that the
xc--kernel has a {\it hyper-non local} behavior and 
memory dependence that is at least one order of magnitude 
stronger than in solids.
The static and 
dynamical polarizabilities are both well described in simple
H$_2$ or in more sophisticated
trans--polyacetylene molecular chains.

The polarizability of low-dimensional systems is proportional 
to the spatial average of the reducible polarization
function, $\chi\(\rr,\rr';\go\)$ that can be obtained from the
solution of the TDDFT equation
\begin{align}
\hat{\chi}\(\go\)=\hat{\chi}_0\(\go\)\[
1+\hat{f}_{Hxc}\(\go\)\hat{\chi}\(\go\)\].
\label{eq:1}
\end{align}
The exchange--correlation part of the kernel $f_{Hxc}=f_{Hartree}+f^{LR}_{xc}$ 
as derived in Refs.~\cite{andrea,sottile}
mimics to a good extent the MBPT results for the polarization function  
based on the solution of the Bethe-Salpeter\,(BS) equation
for the electron--hole dynamics in the basis of the 
G$_0$W$_0$ self--energy quasiparticle states~\cite{rmp}.  
This kernel is
first order in the screened coulomb potential $W$:
$f_{xc}^{LR} = \chi_0^{-1} W \chi_0^{-1}$, where $\chi_0$ is the 
non-interacting quasi--particle response function.
$f_{xc}^{LR}$ includes 
the correct behavior 
$\lim_{\qq\rightarrow 0} f_{xc}^{LR}\(\go,\qq\)\sim-\(\gc+\gb\go^2\)/|\qq|^2$
in the long--range 
regime~\cite{andrea,sottile,botti}.
The two constants, $\gc$ and $\gb$, measure respectively
the degree of spatial non locality
and memory effects.
In spite of its simplicity and 
in contrast to EXX and CDFT, this $f_{xc}^{LR}$
fully captures the excitonic features in the absorption spectra
of bulk semiconductors, insulators as well as surfaces\cite{tddft}. 
\begin{figure}[H]
\begin{center}
\epsfig{figure=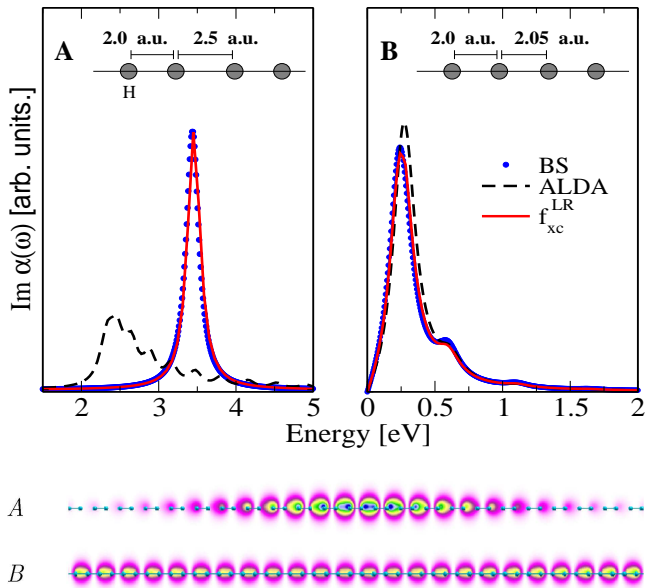,width=8.5cm}
\end{center}
\caption{ {\footnotesize (Color on-line)
Axial optical polarizability spectra of the molecular
H$_2$ chain A, (Peierls distorted, left frame) and chain B 
(nearly equally spaced H-atoms, right frame).
While the present $f_{xc}^{LR}$ kernel yields an excellent 
agreement with the Many--Body calculations in both cases, 
ALDA works only in the small gap case when the exciton is not localized.
The excitonic wavefunctions~\cite{excwf} corresponding to the main absorption peak are
shown fixing the the hole in the middle of the central hydrogen molecular
bond. The delocalisation of the exciton wavefunction for 
the very small gap (chain B) is connected with a successful description of the spectra
by the ALDA (see text).}}
\label{fig:1} 
\end{figure}

In order to illustrate how the kernel $f_{xc}^{LR}$ solves the 
failure of ALDA (and of any semi-local approximations)  
we consider two Peierls distorted
infinite molecular $H_2$ chains~\cite{method}. In chain A
(see Fig.~\ref{fig:1}) the inter--molecular distance is set to 2.5~a.u. and 
the intra-molecular distance is 2.0~a.u.
(as in previous works\cite{vanfaassen,champagne,mori}).
This chain is a semiconductor 
with an LDA gap of 2.28 eV. In chain B the inter-molecular distance
is reduced to 2.05~a.u. and the LDA gap gets very small (0.26 eV).
In Fig.~\ref{fig:1} we compare the calculated TDDFT absorption spectra
for chains A and B in the ALDA and MBPT--$f_{xc}^{LR}$. As the electronic 
density is more homogeneous in chain B than in chain A 
the ALDA gives an almost indistinguishable dynamical
polarizability if compared to the BS calculations (see Fig~\ref{fig:1} right panel). 
For the
chain A, however, the ALDA performance worsens 
considerably, and the main absorption peak is not at all reproduced.
The $f_{xc}^{LR}$ kernel, instead, yields a dynamical polarizability in both
chains almost indistinguishable 
from the BS calculations. 
Similarly to the case of wide gap insulators~\cite{andrea} the correct description
of excitonic states require the $f_{xc}^{LR}$ kernel to have non local 
Fourier components. This means that $f_{xc}^{LR}\(\go,\qq\)$ is a matrix, 
whose size (a few hundred reciprocal space vectors in the present case) is intimately related to the 
localisation of the excitonic state~\cite{andrea}.

The failure of the ALDA can be understood by looking at the spatial dispersion
of the excitonic state corresponding to the main absorption peaks
shown at the bottom of Fig.~\ref{fig:1} for both chains. In 
chain A, where ALDA does not work, the exciton is confined within 
few $H_2$-monomers ($\simeq$ 36 a.u.). Consequently, the excitonic dispersion
introduces a characteristic length given by the exciton linear extension.
In chain B the exciton is basically spread all over the chain.
If we now look at the xc--kernel $f_{xc}^{LR}$ for
the two chains we see very drastic differences: 
whereas in chain B both 
$\gc$ and $\gb$ are almost zero,
in chain A $\gc\sim 16.92$ and $\gb\sim 0.45$ eV$^{-2}$\cite{volume_dep_note}, which
is more than one order of magnitude larger than in solids,
where $\gc\sim 0.1-1.5$, while $\gb\sim 1-34\times 10^{-3}$ eV$^{-2}$~\cite{botti}. 
This large value of $\gb$ reflects the dynamical
 renormalization of the excitonic energy, 
due to memory effects.
More importantly the static limit of the long--range part of the
total TDDFT kernel, 
$\lim_{\go\rightarrow 0} f_{Hxc}\(\go,\qq\)\sim {\(4\pi-\gc\)}/|\qq|^2$ is negative. This means that
the spatial non locality of $f_{xc}^{LR}$ is stronger than the Hartree term~\cite{solid_vs_chain}.
This {\it hyper non locality} and the memory effects in $f_{xc}^{LR}$ 
cannot be captured by  any static, local or semi-local approximation.

\begin{figure}[H]
\begin{center}
\epsfig{figure=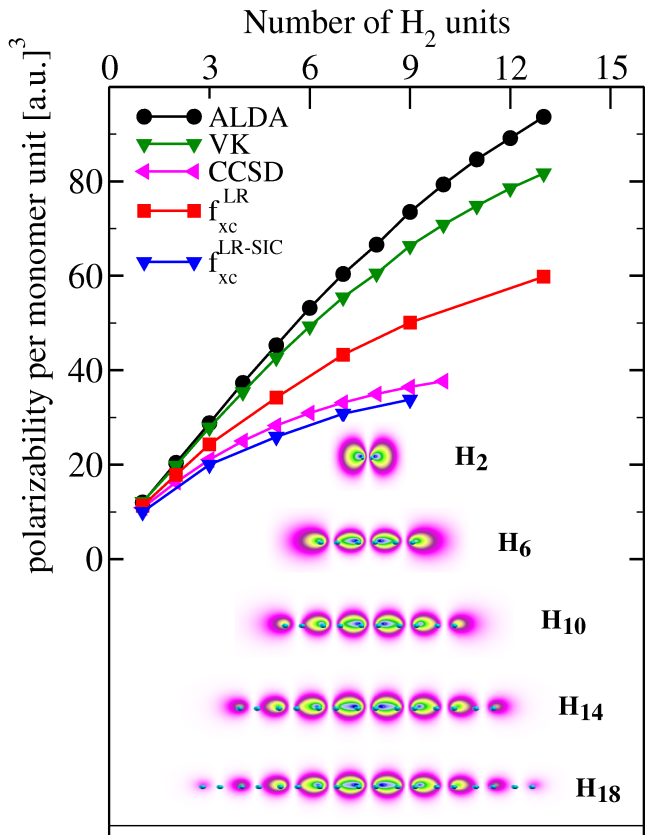,width=8.5cm}
\end{center}
\caption{{\footnotesize (color on-line) Axial polarizability per monomer
of the molecular H$_2$ chains.
TDDFT calculations within the ALDA, $f_{xc}^{LR}$, 
and $f_{xc}^{LR+SIC}$ (see text) are
compared with coupled cluster CCSD(T) results~\cite{champagne}. 
and CDFT using the VK kernel~\cite{vanfaassen}.
The real--space excitonic wavefunctions of selected chains,
with the hole placed in the middle of the central hydrogen molecule bond,
are also showed.}}
\label{fig:2}
\end{figure}

Turning now to finite-size effects, we show in Fig.~\ref{fig:2} the 
results for the static polarizability of 
finite length  H$_2$ chains within different approximations for the
xc--kernel.  Both ALDA and CDFT yields a very slow
saturation of $\ga$ as a function of the chain length and a 
strong overestimation of the static polarizability compared to
accurate quantum chemical coupled cluster results CCSD(T). 
The TDDFT results obtained using the $f_{xc}^{LR}$
kernel partially reduce the
ALDA and CDFT overestimation, showing a faster
saturation. However the agreement with CCSD(T)
is not yet satisfactory.
We have traced the residual discrepancy to the lack of
SIC in the LDA wavefunctions used to construct the xc--kernel. This
can be easily corrected by recomputing the kernel, $f_{xc}^{LR-SIC}$
using Hartree--Fock\,(HF) self--interaction--free
wavefunctions to build $\chi_0$. 
This new kernel yields
axial polarizabilities in excellent agreement with
the CCSD(T) results. The SIC
increases the xc--kernel spatial non locality factor $\gc$,
which translates into further confinement of the excitonic states, compared
to  the LDA-based $f_{xc}^{LR}$ results.
This can be rationalized by looking at the excitonic
wave--functions shown in the bottom of Fig.\ref{fig:2}.
The probability of finding an electron at the end of the chain when the hole 
is placed in the middle of the bond of the central $H_2$ molecule 
(that measures excitonic confinement)
quickly decreases while increasing the chain length, being almost zero for
for $N=9$. The polarizability follows
the same trend, saturating when  the excitonic 
wavefunction dispersion does not change anymore.
We should mention that, for small chains, 
ALDA gives good results as
the excitonic wavefunction is spread over the whole  
molecule, and the electronic homogeneity is not strongly
perturbed (like in the chain B case discussed above).
It is clear that exciton confinement plays a major
role in the determination of the response properties of $H_2$
molecular chains providing microscopic support to the 
relevance of the ultranonlocality concept
in exchange-correlation functionals.

\begin{figure}[H]
\begin{center}
\epsfig{figure=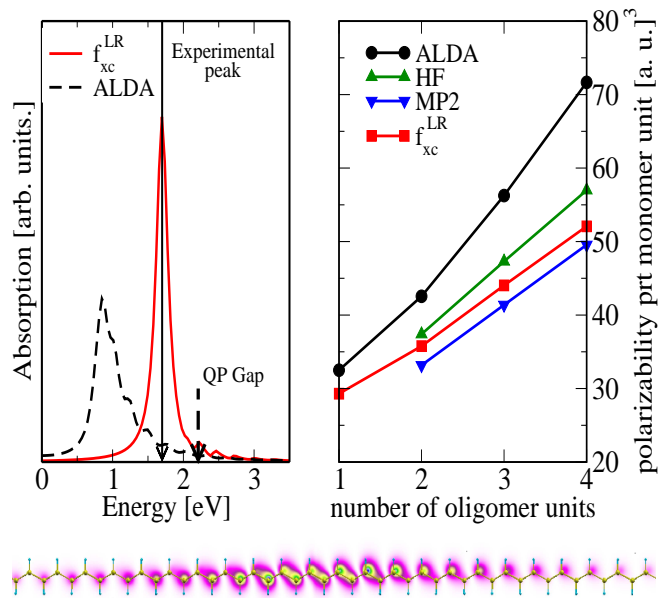,width=.48\textwidth}
\end{center}
\vspace{-.5cm}
\caption{{\footnotesize (color on-line) 
Calculated absorption spectra for the infinite (left frame) and polarizability 
per monomer for the finite (right frame)
trans--polyacetylene chain. ALDA is compared with
the $f_{xc}^{LR}$. 
The quasi-particle gap 
and the position 
of the  experimental absorption at 1.7 eV~\cite{leising}  are also showed.
The axial polarizability
for the finite chains calculated within HF~\cite{kirtman} and
MP2~\cite{toto} are also reported.
In the bottom we show the excitonic wave function corresponding to the bright 
$1.69$\,eV peak of the infinite chain.
}}
\label{fig:3}
\end{figure}

The H$_2$ chain represents an extreme case of system with few electrons, where
confinement is crucial, while electronic screening  is negligible (this is the 
reason why time--dependent HF works fine). For these reason TDDFT kernels
that work in the case of H$_2$ chain may fail in more complex molecular systems
where electronic screening becomes more important. This is the case of 
the trans--polyacetylene\,(PA) molecular chain.
The PA chain has  been extensively studied in the past
within the BS scheme~\cite{rohlfing}. It has been
showed that the main absorption peak observed experimentally is excitonic, with $0.44\,eV$
binding energy. On the contrary all available calculations within TDDFT do not
reproduce the experimental results: ALDA spectra (see Fig.~\ref{fig:3}) is too low
in energy and the main absorption peak shape is more similar to a Van Hove singularity
than to an isolated excitonic peak.
Within an EXX calculation it is possible to
obtain a good agreement with the experiment only by {\it neglecting the x--kernel}, i.e.
in the independent particle approximation~\cite{rohra}, which is clearly
inconsistent.
For finite length chains,
as in polyethylene, EXX results are, again, very similar
to HF~\cite{mori} that, in the case of PA largely overestimates
the quantum chemistry MP2 results (inset of Fig.~\ref{fig:3}).
As shown in Fig.~\ref{fig:3},
the present $f_{xc}^{LR}$ kernel, instead, is in very good agreement
with the MP2 results in the finite length
case and with the experimental absorption
peak position in the infinite length case~\cite{comment}.
MP2 calculations~\cite{toto} predicts the PA chain
polarizability to saturate  around 15 monomers, in excellent agreement with the
excitonic extension obtained with the present $f_{xc}^{LR}$ kernel (shown
in the bottom of Fig.~\ref{fig:2} for the bright PA exciton at 1.69~eV).
As a result even in the more complex PA chain the polarizability saturation
naturally correlates with the excitonic localisation length 
fully confirming the physical picture that emerged from the previous results
for the H$_2$ chains.

In conclusion we have shown that TDDFT successfully explain the optical saturation in
molecular chains in terms of excitonic confinement. The Many-Body based
xc--kernel, with the correct long--range tail and including, in the H$_2$ chains, self--interaction
corrections yields static and dynamical polarizabilities in excellent
agreement with accurate quantum chemical calculations.
We have proved that there exists a close link between the 
excitonic spatial extension and the axial polarizability
in the TDDFT framework, giving a sound interpretation of the severe breakdown 
of the local--density approximation in anisotropic structures.

The authors thank Myrta Gr\"uning for her contributions at
the early stage of this work as well as Rosa Di Felice
for fruitful discussions.
We acknowledge support by the European Community
Network of Excellence Nanoquanta (NMP4-CT-2004-500198),
SANES (NMP4-CT-2006-017310), DNA-NANODEVICES (IST-2006-029192),
and Spanish  MEC (FIS2007-65702-C02-01) projects,
Basque Country University (SGIker ARINA), and
Barcelona supercomputing Center Mare Nostrum.


\begin{thebibliography}{50}
\bibitem{excwf} The excitonic wave--functions have been obtained from the
the linearized  Bethe--Salpeter equation by a diagonalization preocedure.
\bibitem{volume_dep_note} In low--dimensional systems the two constants 
$\gc$ and $\gb$ scale linearly as a function of the simulation volume.
Nevertheless  the $f_{Hxc} \chi_0$ product entering the TDDFT equation is always
finite.
\bibitem{solid_vs_chain}
In solids, the optical absorption $\gee$ differs from the polarizability $\ga$.
$\gee$ is proportional
to the spatial average of the modified reducible polarization $\overline{\chi}$, solution
of Eq.\ref{eq:1} with no long--range component in the Hartree term of the $f_{Hxc}$ 
kernel~\cite{rmp}, i.e.,
$\lim_{\qq\rightarrow 0} f_{Hxc}^{solids}\(\qq\)=f_{xc}\(\qq\)$ and even small
values of $\gc$  yields  large modifications of the non--interacting absorption spectrum.
In the infinite H$_2$ chain $f_{xc}^{LR}$ must counteract the 
non zero Hartree term to yield similar effects on the polarizability spectrum.
\bibitem{comment} In contrast to the $H_2$ molecular chains, 
the LDA wavefunctions do capture very well the delocalised nature of the $\pi$--$\pi^*$ orbitals
in the PA chain. Therefore, additional SIC corrections are less important and
LDA wavefunctions are a good starting point for the MBPT--kernel (as 
in previous MBPT calculations\cite{rohlfing}).
\end{thebibliography}
\end{document}